\documentclass[12pt]{article}
\usepackage{amssymb,amsmath}
\usepackage{natbib}
\usepackage{graphicx}
\usepackage{epsfig}

\begin{document}

\title{Marginal resonances and intermittent behaviour
in the motion in the vicinity of \\ a separatrix}
\author{Ivan~I.~Shevchenko \\
Institute of Theoretical Astronomy, \\
Russian Academy of Sciences, \\
Nab.~Kutuzova~10, St.Petersburg 191187, Russia}

\date{}

\maketitle

\abstract{

A condition upon which sporadic bursts (intermittent behaviour) of
the relative energy become possible is derived for the motion in
the chaotic layer around the separatrix of non-linear resonance.
This is a condition for the existence of a marginal resonance,
i.e. a resonance located at the border of the layer. A separatrix
map in Chirikov's form [Chirikov, B. V., Phys. Reports 52, 263
(1979)] is used to describe the motion. In order to provide a
straightforward comparison with numeric integrations, the
separatrix map is synchronized to the surface of the section
farthest from the saddle point. The condition of intermittency is
applied to clear out the nature of the phenomenon of bursts of the
eccentricity of chaotic asteroidal trajectories in the 3/1 mean
motion commensurability with Jupiter. On the basis of the
condition, a new intermittent regime of resonant asteroidal motion
is predicted and then identified in numeric simulations. }


\section{Introduction}

By the intermittent behaviour of a dynamical system one usually
implies a chaotic one representing a sequence of intervals of
two or more qualitatively different regimes of motion which
alternate in a random fashion. An example is an apparently-chaotic
behaviour with embedded intervals of quasiregular motion; another
example represents a sequence of quasiregular intervals of random
duration, separated by some pronounced regular-looking events
(say, bursts). The theory of intermittency is well developed in the
domain of dissipative dynamical systems, and in particular in the
case of one-dimensional maps~\cite{BPV88}. Hamiltonian intermittency
has yet attracted much less attention, though it is not at all
a rare phenomenon. A short review of it is made by Zaslavsky et al.
in~\cite{ZEA91}. They also consider in detail its mechanism in the
Fermi model of acceleration of particles in regular electromagnetic
fields.

In the paper which follows, a particular case of Hamiltonian
intermittency, characterized by random burst-like events
(i.e. bursts of an action-like variable
separated by quiet quasiregular intervals of random duration),
is studied. Namely, the burst-like intermittency
in the motion in a vicinity of the separatrix of non-linear
resonance is analysed. This is performed in the framework
of a straightforward theory of marginal resonances, i.e. resonances
located at the border of the chaotic layer inside it, or,
in the most prominent appearance, resonances the unperturbed
separatrices of which would be tangent to the layer's border.
As an application, sporadic bursts of the eccentricity
of asteroidal orbits in the 3/1 mean motion commensurability
with Jupiter are considered. To find out conditions for emergence
of such bursts analytically is one of motivations of
the following study.

The primary motivation is to derive a condition
for the intermittency, as defined above, for the motion near
the separatrix of non-linear resonance.
To describe the motion near the separatrix, I~use the separatrix map
in Chirikov's form~\cite{C79}, or, in identical terms,
the ``whisker map''.

\section{The separatrix map}

Consider the Hamiltonian

\begin{equation}
H = H_0 + \Lambda \cdot (\xi_1 + \xi_2),
\label{hp}
\end{equation}

\noindent
where $H_0 = {{{\cal G} p^2} \over 2} - {\cal F} \cos \kappa$
is the pendulum Hamiltonian, and the perturbation,
which is periodic and symmetric, is given by the terms
$\xi_1 = \cos (\kappa - \sigma)$,
$\xi_2 = \cos (\kappa + \sigma)$, where
$\sigma = \Omega t + \sigma_0$.
The angle $\kappa$ is the pendulum's angle, and $\sigma$ is
the phase angle of perturbation;
$\sigma_0$ is the initial phase; $p$ is the momentum;
$\Omega$ is the constant perturbation frequency;
${\cal F}$, ${\cal G}$, $\Lambda$ are constants.

According to Chirikov~\cite{C79},
the motion of the system with Hamiltonian~(\ref{hp})
in a vicinity of the separatrix is described by the
separatrix map (hereafter SM):

\begin{eqnarray}
w_{n+1} & = & w_n - W \sin \sigma_n,  \nonumber \\
\sigma_{n+1} & = & \sigma_n +
                   \lambda \ln {32 \over \vert w_{n+1} \vert}
                   \ \ \ (\mbox{mod } 2 \pi),
\label{sm}
\end{eqnarray}

\noindent
where $w$ denotes the pendulum's relative energy:
$w = {H_0 \over {\cal F}} - 1$.
Constants $\lambda$ and $W$ are
parameters: $\lambda$ is the ratio of
$\Omega$, the perturbation frequency, to
$\omega_0 = ({\cal F G})^{1/2}$,
the frequency of small-amplitude pendulum oscillations; and
$W$ is given by the formula

\begin{equation}
W = {\Lambda \over {\cal F}} \lambda
\left[ A_2(\lambda) + A_2(-\lambda) \right] =
{\Lambda \over {\cal F}}
{4 \pi \lambda^2 \over \sinh {\pi \lambda \over 2}}.
\label{W}
\end{equation}

\noindent Here

$$ A_2(\lambda) = 4 \pi \lambda {\exp({{\pi \lambda} / 2})
\over \sinh (\pi \lambda)} $$

\noindent is the value of the Melnikov--Arnold
integral as defined in~\cite{C79}. Formula~(\ref{W})
differs from that given in~\cite{C79,LL92} by
the term $A_2(-\lambda)$, which is small for $\lambda \gg 1$
and is usually ignored. However, its contribution is significant
for $\lambda$ small, i.e. in the case of low-frequency
perturbation.

One iteration of the SM corresponds to one
period of pendulum's rotation, or a half-period of its libration.

The motion of system~(\ref{hp}) is mapped by Eq.~(\ref{sm})
asynchronously: the action-like varible $w$ is taken
at $\kappa = \pm \pi$, while the phase angle $\sigma$ is taken
at $\kappa = 0$. Below it will be demonstrated that an
asymmetry of SM-produced phase portraits in relation to
the vertical lines $\sigma = 0$ or $\sigma = \pi$
(see e.g. Fig.~1 in~\cite{C90} or Fig.~6 in~\cite{A95})
is an artifact of the desynchronization.

The desynchronization can be removed by using variables
$\widetilde w_n$, $\sigma_n$, where

\begin{eqnarray}
\widetilde w_n & = & w_n - {W \over 2} \sin \sigma_n - \delta(\lambda) \,
                        W \cos \sigma_n \ =
                                                   \nonumber \\
           & = & {w_n + w_{n+1} \over 2} - \delta(\lambda) \,
                        W \cos \sigma_n,
\label{ws}
\end{eqnarray}

\noindent
instead of $w_n$, $\sigma_n$. This transformation synchronizes
the SM to the surface of section $\kappa = 0$.
Eq.~(\ref{ws}) can be rigorously derived by calculation
of the relative energy increment on the half-period of the map.
If one sets $\delta(\lambda) = 0$, it can be understood as an
interpolation of the two values of $w$ to the surface $\kappa = 0$.
Derivation of $\delta(\lambda)$ constitutes a technical problem
of analytical evaluation of definite integrals,
which is lengthy. Besides, the contribution of the term
with $\delta(\lambda)$ is small for the value of $\lambda$
used below in Section~\ref{ast}, and it is not crucial for the
purposes of this paper. Therefore, a complete derivation
of $\delta(\lambda)$ will be given elsewhere.
Here I present only the final formula

\begin{equation}
\delta(\lambda) = {1 \over \pi}
\left\{ \mbox{Re} \left[ \psi \left({i \lambda \over 2} \right) -
\psi \left({i \lambda \over 4} \right) \right]
+ {1 \over \lambda^2} - \ln 2 \right\}
\sinh {\pi \lambda \over 2},
\label{dlt}
\end{equation}

\noindent
where $\psi(z) = {\Gamma^\prime (z) \over \Gamma(z)}$ is the
digamma function, $i$ is the imaginary unit.
According to e.g.~\cite{AS70},
the real part of the digamma function of an imaginary
argument is given by the series

\begin{equation}
\mbox{Re } \psi (i y) = - C + y^2
           \sum_{n=1}^\infty {1 \over n (n^2 + y^2)},
\label{Repsi}
\end{equation}

\noindent
where $y$ is any real number,
$C \approx 0.577216$ is Euler's constant.

The quantity $\delta(\lambda)$ goes to accordingly plus and minus
infinity when $\lambda \to 0$ and $\lambda \to + \infty$, i.e. the
term with $\delta(\lambda)$ in Eq.~(\ref{ws}) is especially important
to take into account in cases of low and high frequency perturbation.
However, at $\lambda \sim 1 \div 3$ it is close to zero.

I~use the procedure of synchronization to the unified surface
of section below, when comparing phase portraits obtained by
direct numeric integrations with those obtained with the map.

Note that a separatrix map ``shifted'' to a surface of section
with the pendulum's angle taken near the saddle point (i.e. at
$\kappa \approx \pm \pi$ in our terms) was recently derived
by Abdullaev and Zaslavsky~\cite{AZ95,AZ96}
in connection with studies of the behaviour of plasma in tokamaks.
The ``shifting'' of an SM to $\kappa \approx \pm \pi$ requires
a calculation of an increment of the phase angle of perturbation,
instead of the calculation of the increment of the relative
energy in the case of synchronization to the surface
$\kappa = 0$ considered above. For our purposes, incorporating
a comparison with numeric integrations of the motion,
the choice of the latter section is preferable.
Indeed, sections near the saddle point
do not necessarily intercept all trajectories of the libration
mode (those trajectories simply do not reach the value of
$\kappa = \pm \pi$), while the section taken at $\kappa = 0$
provides a complete coverage of the phase space.

Apart from the problem of synchronization, the ordinary SM~(\ref{sm})
has another shortcoming which limits its applicability for modelling
the real motion. Namely, it is valid for a low strength
of perturbation, i.e. for $W \ll 1$. To a certain degree,
this shortcoming can be removed by some complication
of the map: if the perturbation is not weak, it is straightforward,
in order to improve the performance of the map, to replace the
logarithmic approximation of the increment of the
phase angle by its exact value, which depends
on what side of the line of the unperturbed separatrix
the motion takes place. Making this replacement, one has
the exact SM:

\begin{eqnarray}
w_{n+1} & = & w_n - W \sin \sigma_n,  \nonumber \\
\sigma_{n+1} & = & \sigma_n + \Delta \sigma_n
                   \ \ \ (\mbox{mod } 2 \pi),
\label{esm}
\end{eqnarray}

\noindent
where


\begin{equation}
\Delta \sigma_n =
\begin{cases}
 2 \lambda K \left[ \left(1+{w_{n+1} \over 2} \right)^{1/2} \right], &
                                 \mathrm{if} w_{n+1} < 0 ; \cr
 2 \lambda \left(1+{w_{n+1} \over 2} \right)^{-1/2}
          K \left[ \left(1+{w_{n+1} \over 2} \right)^{-1/2} \right], &
                                 \mathrm{if} w_{n+1} > 0 ;
\end{cases}
\nonumber
\label{dsigma}
\end{equation}

\noindent
$K(k)$ is the complete elliptic integral of the first kind
with modulus $k$.

\section{A condition for intermittency}

An opportunity of sporadic strong rapid variations of the
relative energy $w$ in the chaotic motion near the separatrix
depends on the structure of the border of the chaotic layer.
Excursions to high values of the relative energy $w$ become possible,
when, with the increase of the perturbation, the border of the
chaotic layer touches the separatrix of an integer resonance,
and the very narrow chaotic layer around this latter separatrix
is no more separate from the main chaotic layer.
They start to overlap, i.e. a heteroclinic connection emerges
between them.
Thus a marginal integer resonance is formed. In case of half-integer
or other fractional resonances the maximum relative variation of $w$
is much less. Therefore solely the integer case is considered
in what follows.

Since for small perturbations the dynamical behaviour described
by the SM is symmetric with respect to the sign of $w$ (i.e. it is
the same for librations and circulations, see Eq.~(\ref{sm})),
let $w$ be positive. The condition for the tangency is then
$w_{bord} = w^{(m)} - \Delta w^{(m)}$, where $w_{bord}$ is the location
of the border of the main chaotic layer, $w^{(m)}$ is that of the
center of the marginal resonance of an integer order $m \geq 1$, and
$\Delta w^{(m)}$ is the half-width of the marginal resonance.
It is known that $w_{bord} \approx \lambda W$~\cite{C90,CS84} and
$w^{(m)} = 32 \exp({- 2 \pi m / \lambda})$~\cite{LL92}. The quantity
$\Delta w^{(m)}$ can be derived by means of linearization of the
SM with respect to the action-like variable $w$
in the neighbourhood of the fixed point. From consideration of
the Hamiltonian of the inferred standard map one has
$\Delta w^{(m)} = 2 ({w^{(m)} W / \lambda})^{1/2}$.

Finally, the condition for the tangency of the unperturbed
separatrix of a marginal integer resonance to the border of
the main chaotic layer, i.e. the condition for appearance of
bursts, is

\begin{equation}
W = W_t^{(m)}(\lambda),
\label{tc}
\end{equation}

\noindent
where

\begin{equation}
W_t^{(m)}(\lambda) = {32 \over \lambda^3}
    \left[ \left( 1 + \lambda^2 \right)^{1/2} - 1 \right]^2
    \exp \left( - {2 \pi m \over \lambda} \right) .
\label{Wt}
\end{equation}

\noindent
In this approximation, the tangency condition is one and the
same for $w$ both positive and negative, i.e. for both sides
of the main chaotic layer.

Of course, the bursts of the relative energy also take place
at $W$ somewhat greater than $W_t^{(m)}(\lambda)$, but on
increasing $W$ the chaotic layer around the separatrix
of the marginal resonance becomes thicker and thicker,
and the resonance deepens into the main chaotic layer. This
leads to a lesser value of the relative amplitude of bursts,
and also to a loss of their self-similarity.
The most pronounced and self-similar
bursts take place at the edge of the tangency, when
$W = W_t^{(m)}(\lambda)$, or when $W$ is slightly greater
than this value.

The term ``tangency condition'' is used in what follows to denote
the condition for intermittency, though of course
this implies an approximate description of the phenomenon
of the heteroclinic connection in this problem.
This approximation is possible when the chaotic layer
around the separatrix of a marginal resonance is narrow
in comparison with the width of the main chaotic layer.

The extremum value of the relative energy during a burst,
when $w$ e.g. is positive, is $w_{extr} = w^{(m)} + \Delta w^{(m)}$,
or $w_{extr} = w_{bord} + 2 \Delta w^{(m)}$,
or $w_{extr} = 2 w^{(m)} - w_{bord}$. Let us take the latter expression,
since it is the simplest one. For both possible signs of $w$,
the extremum value is finally

\begin{equation}
w_{extr} = \pm \left[ 64 \exp \left( - {2 \pi m \over \lambda} \right)
              - \lambda W_t^{(m)}(\lambda) \right].
\label{wextr}
\end{equation}

What is the numeric precision of the tangency condition~(\ref{tc})
in predicting the onset of intermittency?
This problem is twofold: how good the condition is for describing
the separatrix map behaviour, and how good it is for describing
the motion of an underlying Hamiltonian system (say~(\ref{hp})),
represented by the SM. Concerning the first part of the problem,
it is simplified by an invariance property of SM~(\ref{sm}),
which is made evident by transforming Eq.~(\ref{sm}) to the form
(used e.g. in~\cite{CS84}):

\begin{eqnarray}
     y_{n+1} &=& y_n + \sin x_n, \nonumber \\
     x_{n+1} &=& x_n - \lambda \ln \vert y_{n+1} \vert + c
                   \ \ \ (\mbox{mod } 2 \pi),
\label{sm1}
\end{eqnarray}

\noindent
where $y = {w \over W}$, $x = \sigma + \pi$; and the new parameter is

\begin{equation}
c = \lambda \ln {32 \over \vert W \vert}.
\label{c}
\end{equation}

\noindent
The invariance property is that the dynamical behaviour of
the SM with the parameter $c$ is identical to that with the
parameter $c + 2 \pi k$, where $k$ is any integer.
So, the value of the ``perturbation parameter'' $W$ in the case
of SM~(\ref{sm}) does not have any physical limits, if the SM
is considered separately from an underlying dynamical system.
This inference is not true for exact SM~(\ref{esm}), where
$\vert W \vert$ should be small enough, as considered below.

The accuracy of the tangency condition~(\ref{tc}) is determined by
several factors: by the accuracy of the estimate of location of the
border of the main chaotic layer, by that of location of the center
of a marginal resonance, by that of the width of the latter,
also by not taking into account the necessity of synchronization
of the SM. Therefore, it can be studied only numerically. Since
two parameters are involved, $\lambda$ and $W$, a detailed study
would take much space.
Here I present results on the accuracy of the tangency condition
for the marginal resonance $m=1$, computed for six representative values
of $\lambda$ in the range $0.1 \leq \lambda \leq 10$.
Concerning behaviour of the SM itself, the value of $m$ does not matter.
In representing underlying systems, the accuracy is higher for $m > 1$,
since the strength of perturbation is lower.

The results are accumulated in Table~1.
First, there are values of $c$ at which the onset of intermittency
associated with an integer marginal resonance is observed in the
behaviour of SM~(\ref{sm1}) for a given value of $\lambda$.
On decreasing $c$ (increasing $W$) this onset is signalled by
a sudden jump in the maximum value of $\vert y \vert$ available by
the motion inside the chaotic layer. The threshold value of $c$
for $m=1$ is denoted by $c^{(1)}_t(\lambda)$. Numerically it is
obtained as follows: on varying (decreasing) the parameter $c$
on an interval of length $2 \pi$ with a step equal to $0.01$,
the onset of intermittency with the maximum jump in $\vert y \vert$
is fixed. Runs of length $n_{it}=10^4$ and $n_{it}=10^7$, where
$n_{it}$ is the number of iterations of the map, were performed
for each step in $c$. The threshold values are listed in Table~1.
The theoretical value of $c^{(1)}_t(\lambda)$ is connected to
$W^{(1)}_t(\lambda)$ by Eq.~(\ref{c}).
$\Delta c^{(1)}_t(\lambda)$ is the difference between
$c^{(1)}_t(\lambda)$ observed for $n_{it}=10^7$ and
$c^{(1)}_t(\lambda)$ theoretical.
The value of $W^{(1)}_t(\lambda)$ theoretical is given by Eq.~(\ref{Wt}).
The strength of perturbation $\varepsilon \equiv \Lambda / {\cal F}$
for underlying system~(\ref{hp}) is found by means of Eq.~(\ref{W}).
Substituting $W = W^{(1)}_t(\lambda)$ theoretical in this equation,
one has the threshold value $\varepsilon^{(1)}_t(\lambda)$ for
the underlying system.

\begin{table*}
\begin{center}
\caption{Accuracy of the condition for intermittency}
\begin{tabular}{crrrrll} \hline
\noalign{\smallskip}
          $\lambda$&
          ${c^{(1)}_t(\lambda), \atop n_{it}=10^4}$&
          ${c^{(1)}_t(\lambda), \atop n_{it}=10^7}$&
          ${c^{(1)}_t(\lambda), \atop {\mathrm{theor.}}}$&
          $\Delta c^{(1)}_t(\lambda)$&
          ${W^{(1)}_t(\lambda), \atop \mathrm{theor.}}$&
          ${\varepsilon^{(1)}_t(\lambda), \atop \mathrm{theor.}}$ \\
\noalign{\smallskip}
\hline\\
 0.1& 6.58& 6.58& 6.65&$-0.07$&$4.11 \cdot 10^{-28}$&$5.15 \cdot 10^{-28}$ \\
 0.5& 7.18& 7.21& 7.38&$-0.17$&$1.24 \cdot 10^{-5}$ &$3.44 \cdot 10^{-6}$  \\
   1& 7.83& 7.93& 8.05&$-0.12$&     0.0103          &$1.88 \cdot 10^{-3}$  \\
   2& 9.45& 9.66& 9.59&  0.07 &     0.264           & 0.0607               \\
   5&16.03&16.74&16.32&  0.42 &     1.22            & 5.02                 \\
  10&30.32&31.90&31.31&  0.59 &     1.40            &$3.69 \cdot 10^{3}$   \\
\noalign{\smallskip}
\hline
\end{tabular}
\end{center}
\label{Table1}
\end{table*}

An inspection of Table~1 shows that the performance of the
tangency condition is rather good for $\lambda \leq 2$.
The predicted threshold values of $c$ do not differ much
from real ones. For $\lambda$ equal to 5 and 10, the values of
$\vert \Delta c^{(1)}_t(\lambda) \vert$ are much greater.
The reason is that for higher values of $\lambda$
the border of the main chaotic layer becomes ``blurred''.
Indeed, the value of the stochasticity parameter $K$ of the
standard map locally approximating the separatrix map~(\ref{sm1})
at distance $y$ from the line of the unperturbed separatrix $y=0$
is $K = \lambda/y$~\cite{C79}. Hence, the gradient of $K$ with
$\vert y \vert$ in the vicinity of the border
(where $\vert y \vert \approx \vert w_{bord} / W \vert \approx \lambda$,
see above) is $\approx - 1/\lambda$ and is therefore small when
$\lambda$ is high. In other words, the border is not well-defined at
high values of $\lambda$. Therefore, threshold values of $c$
are also not well-defined when $\lambda$ is high.

Concluding on the first part of the problem of the accuracy
of the tangency condition, note that when solely the behaviour
of SM~(\ref{sm1}) or, equivalently,~(\ref{sm}) is concerned,
the tangency condition is applicable to marginal resonances of
arbitrary integer order, i.e. $m$ zero and negative as well.
The threshold value of $c$ for order $m$ is simply related
to that for order $1$:

\begin{equation}
c^{(m)}_t(\lambda) = c^{(1)}_t(\lambda) + 2 \pi (m-1).
\label{cm}
\end{equation}

\noindent
There is no analogous property for exact SM~(\ref{esm}) and for
underlying systems. The tangency condition can be applied to them only
when $m$ is at least greater than zero, since the formula for location
of the center of an integer resonance
$w^{(m)} = \pm 32 \exp({- 2 \pi m / \lambda})$~\cite{LL92}
for $m=0$ already gives $w^{(m)} = \pm 32$, and the relative energy of
a pendulum has the physical limit $w \geq -2$.

The last two columns of Table~1 deal with the second part of the
accuracy problem, i.e. the part concerning relevance to the behaviour
of underlying systems. Generally speaking, here the discrepancies
may be of course greater;
e.g. a study by Veerman and Holmes~\cite{VH86} of a pair of linearly
coupled pendula shows that real widths of resonances
may strongly deviate from theoretical estimates already at rather
low values of the strength of perturbation, long before the resonances
start to overlap.
Besides, the accuracy of the tangency condition in application
to underlying systems is determined by the yet largely non-studied
performance of the SM itself in describing the real motion.
Formula~(\ref{W}) for the parameter $W$ of
the SM is valid for a specific form of perturbation, given by the
functions $\xi$ above; but the SM~(\ref{sm}) itself is far more general.
It can describe motion for other kinds of perturbation as well.
So, even if one studied in full the accuracy of the tangency condition
in the case of the form of perturbation given above,
this would not provide a complete picture.

However, this case is itself of general interest. The strength of
perturbation $\varepsilon^{(1)}_t(\lambda)$ for this case, as noted
already, is calculated as equal to $\Lambda / {\cal F}$ in Eq.~(\ref{W}),
where $W$ is set to be equal to $W^{(1)}_t(\lambda)$ theoretical
given by Eq.~(\ref{Wt}).
An inspection of Table~1 shows that the values of $W^{(1)}_t(\lambda)$
and $\varepsilon^{(1)}_t(\lambda)$ both are very low at $\lambda < 2$.
Author's numeric experience obtained in the wider range
$0.1 \leq \lambda \leq 10$ shows that the synchronized
exact SM~(\ref{esm}) provides authentic phase portraits of the
behaviour of underlying system~(\ref{hp}) already at values of
$\varepsilon$ as large as $0.01$ and even greater;
therefore the accuracy of the tangency condition in the range
$0.1 \leq \lambda < 2$ should be approximately the same for
system~(\ref{hp}) as it is for the separatrix map. The synchronized
phase portrait provided by exact SM~(\ref{esm}) in the case of
$\lambda=2$ and $W = W^{(1)}_t(2)$ is also in good accordance
with the computed behaviour of system~(\ref{hp}), though
$W^{(1)}_t(2)$ and $\varepsilon^{(1)}_t(2)$ are rather high.
Concerning cases $\lambda = 5$ and $10$, the values of
$W^{(1)}_t(\lambda)$ and consequently $\varepsilon^{(1)}_t(\lambda)$
predicted for them (see Table~1) are completely
irrelevant; the motion cannot be even modelled by exact
SM~(\ref{esm}), when the value of $W$ is so large. In order that
$\varepsilon^{(1)}_t(5)$ and $\varepsilon^{(1)}_t(10)$
were less than say $0.01$ (then $W^{(1)}_t(5)$ and $W^{(1)}_t(10)$
are also less, and much less, than $0.01$), the order $m$ of
marginal resonance should be greater than $5$ and $21$ for these
two values of $\lambda$ respectively.
In other words, greater the value of $\lambda$, greater is the
order of the marginal resonance that can be described by the SM.
Remember however that due to the blurring of the border of the
main chaotic layer, the tangency condition becomes irrelevant at
such high values of $\lambda$ in any case, though the SM itself may
describe the motion perfectly.

The conclusion on the accuracy of the tangency condition~(\ref{tc})
in the studied range $0.1 \leq \lambda \leq 10$ is as follows:
for $\lambda$ approximately less than $2$, the condition adequately
predicts onsets of intermittency in the behaviour of the SM
and most likely in that of underlying systems; for greater values
of $\lambda$, the accuracy is much less due to the blurring of the
border of the main chaotic layer. The conclusion is valid for all
integer $m$ (order of marginal resonance) in the case of
SM~(\ref{sm}) or~(\ref{sm1}), and for $m \geq 1$ in the cases of
exact SM~(\ref{esm}) and underlying systems.

Now let us see how our formulae work when possibility of
intermittent behaviour should be identified in a Hamiltonian system
more complicated than that given by Eq.~(\ref{hp}).

\section{Intermittency in the chaotic motion
of asteroids in the 3/1 Jovian resonance}
\label{ast}

In the planar-elliptic Sun--Jupiter--asteroid problem,
when the 3/1 mean motion commensurability is present,
behaviour of certain chaotic asteroidal trajectories
is known to be intermittent: they display multiple eccentricity
bursts randomly spaced in time.
An example of such an orbit with bursts of one and the same
regular shape, and quiet intervals between them of random
duration, is presented by Wisdom in Fig.~13 of his
paper~\cite{W83}. A burst in eccentricity is nothing but
a transition to a particular resonance between variations of
angle coordinates of an asteroid. Transitions to such resonances
display a diversity of behaviour (see e.g.~\cite{FM93} and
references therein).

Shevchenko and Scholl~\cite{SS96} showed that the statistical
distribution of duration of intervals between eccentricity bursts
for intermittent orbits in the 3/1 Jovian resonance
represents, in the tail of the distribution, an algebraic decay.
This is explained by sticking of orbits to the chaos border,
and is similar to the distribution law found
by Chirikov and Shepelyansky~\cite{CS84}
for durations of Poincar\'e recurrences
in computations with the SM.

By means of an analysis of spectra of winding numbers of the
chaotic asteroidal motion in the 3/1 Jovian resonance,
Shevchenko~\cite{S96} found out that
a certain sticking regime of the intermittent motion
can be approximated by the SM, the pendulum's angle
being equal to $\kappa = \widetilde \omega + \varphi$ and the phase angle
of perturbation being equal to $\sigma = 2 \widetilde \omega + \varphi$,
where $\varphi = l - 3 l_J$, and $l$, $l_J$ are the mean longitudes
of an asteroid and Jupiter,
$\widetilde \omega$ is the longitude of the asteroid's perihelion.
This approximation is valid when the motion takes place at the
circulation side of the main chaotic layer (i.e. where the
$\kappa$ angle rotates) near its border.

It cannot be used on the libration side. The paradigm
for the whisker map demands that $\sigma$ should circulate everywhere,
and with constant frequency. But an empirical fact is that when
$\kappa$ librates (e.g. during eccentricity bursts), $\sigma$ also
librates. Thus the paradigm fails. What is more,
the whisker map~(\ref{sm}) cannot be used on this side (even if
the $\sigma$ angle circulated here) because the perturbation terms
in the Hamiltonian of the asteroidal problem (see e.g.~\cite{S96})
do not represent pairs of symmetric counterparts
(such as $\cos (\kappa - \sigma)$ and $\cos (\kappa + \sigma)$
in the paradigm~(\ref{hp})).
The whisker map cannot be used also on the circulation side
in a vicinity of the central line of the main chaotic layer, because
the phase angle $\sigma$ rotates in a regular fashion only during
Poincar\'e recurrences of long enough duration (say, $N > 10$,
in the number of iterations of the approximating map).
So, it can be used only to model the motion on the circulation side
near the border of the main chaotic layer. However this is
sufficient for a study of the intermittent behaviour,
since the latter is determined by the resonant structure of the
chaos border, namely by existence of marginal resonances.

Resonant structure of the motion in a vicinity of the chaos border
can be investigated by means of construction of
spectra of winding numbers (SWN). A version of such a spectrum
is built, according to~\cite{S96}, as follows.
For every Poincar\'e recurrence, its duration $N$
(in the map's iterations) versus its winding number $Q$ (= total
increment in the phase angle $\sigma$, devided by $2 \pi N$) is
plotted. Sticking to resonances results in peaks in the spectrum.

Analogous spectra built for asteroidal orbits in the problem
under study usually have peaks at $Q = 5/4$ and $2$ (only rotation
side of the layer is taken into account).
The details on the spectra see in~\cite{S96}.

\begin{figure}
\begin{center}
\includegraphics[width=4cm,angle=270]{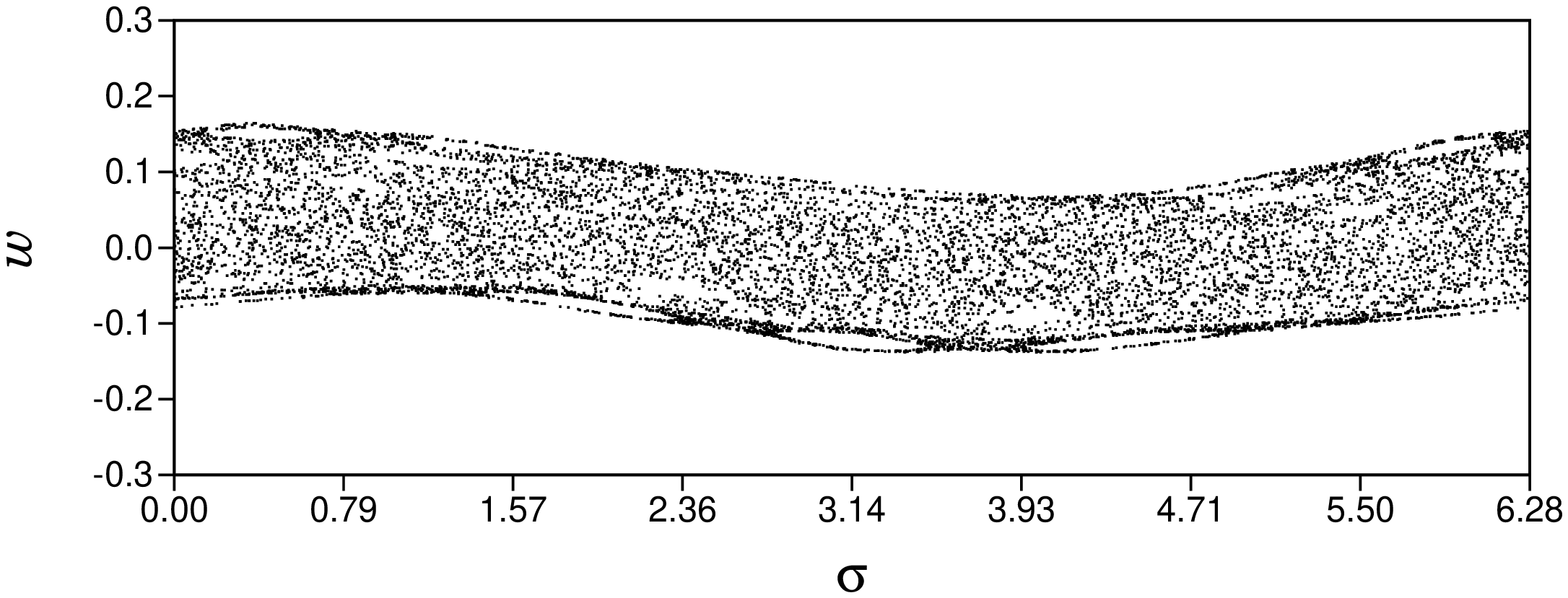}
\end{center}
\caption{The phase plane of the exact SM with $\lambda = 1.34$,
$W = 0.052$.}
\label{ESM1}
\end{figure}

The values of parameters of the approximating SM,
found by means of construction of the SWN for a particular
typical trajectory in the 3/1 Jovian resonance, the starting data
for which are given below, are $\lambda = 1.34$, $W = 0.052$
(i.e. $c = \lambda \ln (32/ \vert W \vert) = 8.6$, cf.~\cite{S96}).
The phase portrait, computed with the exact SM (Eq.~(\ref{esm}))
with these values of parameters, is shown in Fig.~\ref{ESM1}.
A quiet behaviour without bursts is observed at both sides of
the main chaotic layer.

Let us calculate the tangency condition~(\ref{tc}) for the
inferred value of $\lambda$, when the marginal resonance has
the order $m = 1$. From Eq.~(\ref{Wt}) one has
$W_t^{(1)}(1.34) = 0.055$.
The extremum value of the relative energy, given
by Eq.~(\ref{wextr}), is $w_{extr} = \pm 0.51$.
These estimates agree with the real
behaviour of the SM.
In computations with the exact SM (Eq.~(\ref{esm}),
the number of iterations $n_{it}$ equal to $10^7$
was used for a run), when the perturbation is gradually increased,
the bursts at the circulation side appear
at $W=0.053$. At this moment, $w_{extr} = 0.50$.
At the libration side,
the bursts appear somewhat later, at $W=0.063$, with
the amplitude $w_{extr} = -0.63$.

\begin{figure}
\begin{center}
\includegraphics[width=4cm,angle=270]{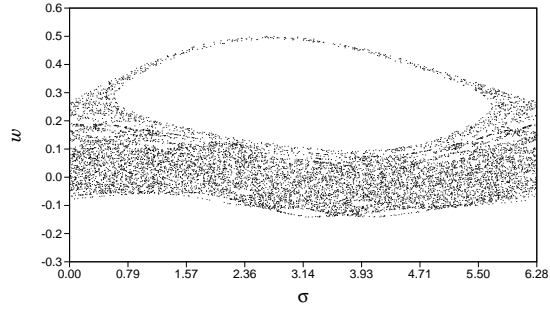}
\end{center}
\caption{The same as in Fig.~\protect\ref{ESM1},
except that $W = 0.055$.
The perturbation is slightly increased, and burst-like excursions
to high values of the relative energy become possible.}
\label{ESM2}
\end{figure}

For the values of the SM parameters
predicted by the tangency condition,
the structure of the chaotic layer is shown in Fig.~\ref{ESM2}.
A thin loop emerges at the layer's border.
Due to sporadic excursions to this loop, the action-like variable
exhibits sequences of high-amplitude bursts.

\begin{figure}
\begin{center}
\includegraphics[width=4cm,angle=270]{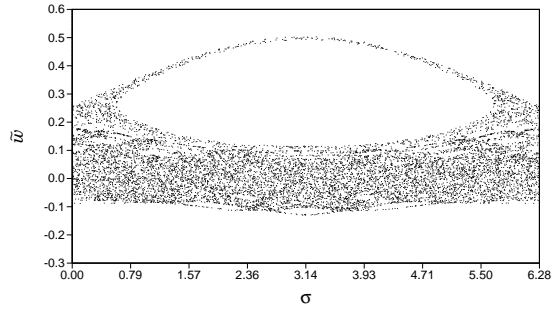}
\end{center}
\caption{The same as in Fig.~\protect\ref{ESM2}, but synchronized
to the unified section $\kappa = 0$.}
\label{ESM3}
\end{figure}

In Fig.~\ref{ESM3}, the synchronized phase portrait is built
at the unified surface of section $\kappa = 0$, by means
of the procedure of synchronization of the SM, as described by
Eqs.~(\ref{ws}, \ref{dlt}) (one has $\delta(1.34) \approx 0.173$).
The asymmetry with respect to the line $\sigma = \pi$, prominent
in Fig.~\ref{ESM2}, disappears in Fig.~\ref{ESM3}. Thus this
asymmetry is an artifact of desynchronization of the
original SM.

\begin{figure}
\begin{center}
\includegraphics[width=4cm,angle=270]{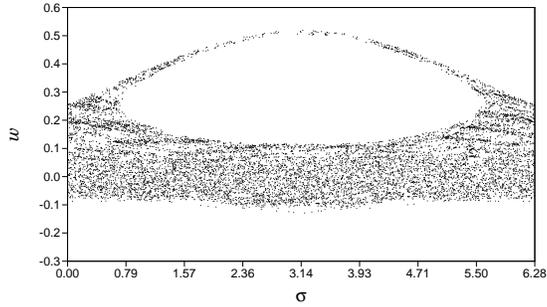}
\end{center}
\caption{A direct numeric integration of the system modelled above by
the exact SM in Figs.~\protect\ref{ESM2} and~\protect\ref{ESM3}.}
\label{DI}
\end{figure}

In Fig.~\ref{DI}, the same phase plane as in Fig.~\ref{ESM3},
but obtained by a direct numeric integration
of the system with Hamiltonian~(\ref{hp}), is presented.
The values of parameters in Eq.~(\ref{hp}) correspond to
the values $\lambda = 1.34$ and $W = 0.055$ used in the exact SM;
namely, ${\cal F} = 1$, ${\cal G} = 2$, $\Lambda = 0.009853$
and $\Omega = 1.34 \sqrt{2}$.
The integration was performed by the 8th order
Dormand--Prince technique~\cite{HNW87} with stepsize control.
The local tolerance was set to $10^{-10}$.

The computed values of variables $w$ and $\sigma$ were taken
straightforwardly at the unified surface of section $\kappa = 0$;
thus the phase plane $w$, $\sigma$ in Fig.~\ref{DI} can be directly
compared to the ``synchronized'' phase plane $\widetilde w$, $\sigma$
of the separatrix map in Fig.~\ref{ESM3}.
One can see that the real system's behaviour (Fig.~\ref{DI})
is adequately described by the phase portrait of the synchronized
exact SM (Fig.~\ref{ESM3}).

The stated values $\lambda = 1.34$, $W = 0.052$ of the SM
parameters were derived in~\cite{S96} for an intermittent
asteroidal trajectory which is in the 3/1 mean motion
commensurability with Jupiter. The orbit was computed
by means of Wisdom's map~\cite{W83} in the planar-elliptic
Sun--Jupiter--asteroid problem. The starting values are
as follows: the mean longitude $l_0 = \pi$,
the longitude of perihelion $\widetilde \omega_0 = 0$,
the semimajor axis $a_0 = 0.4806$, and the eccentricity
$e_0 = 0.05$. Jupiter's eccentricity $e_J = 0.048$, the
longitude of Jupiter's perihelion is zero.
Wisdom~\cite{W83} gives a plot of eccentricity versus time
for this trajectory (see Fig.~13 in his paper),
which displays multiple bursts.

The values $\lambda = 1.34$, $W = 0.052$ derived
for this intermittent asteroidal trajectory, being
compared with the tangency condition
$W = W_t^{(1)}(1.34) = 0.055$, tells one that these values
are at the edge of emergence of bursts of the action-like
variable at the circulation side of the main chaotic layer,
i.e. at the side where the $\kappa$ angle rotates.
In case the perturbation is slightly increased, and the
tangency condition~(\ref{tc}) starts to hold, there should emerge
an intermittent regime in the asteroidal motion. This regime
corresponds to transitions to the motion in the marginal resonance
$m=1$ at the circulation side of the main chaotic layer.
It is different from the well-known regime of major eccentricity
bursts, since the latter takes place at the $\kappa$-libration side.

It turns out that such a regime indeed emerges, but is rare.
Its presence can be identified by means of construction
of a spectrum of winding numbers, as described above.
Construction of the SWN allows one to visualize
the resonant structure of motion~\cite{S96}.
The stated asteroidal orbit, when it sticks to the border of
the main chaotic layer at the circulation side,
has the asymptotic winding number $Q=5/4$~\cite{S96}.
The winding number is defined as the ratio of
rotation frequencies of angles $\sigma$ and $\kappa$.
This corresponds to the definition of $Q$ for the
approximating SM. If, in a case of a particular asteroidal orbit,
the tangency condition~(\ref{tc}) with $m=1$ holds,
the regime of the motion in the marginal resonance has $Q=1$.
The regime manifests itself in the SWN by a peak
at the winding number $Q=1$.

\begin{figure}
\begin{center}
\includegraphics[width=4cm,angle=270]{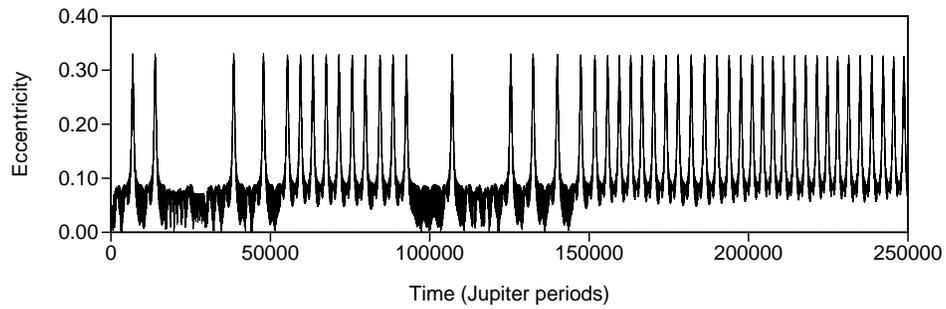}
\end{center}
\caption{The ``usual'' behaviour of the eccentricity of
an intermittent asteroidal trajectory.}
\label{ECC}
\end{figure}

\begin{figure}
\begin{center}
\includegraphics[width=4cm,angle=270]{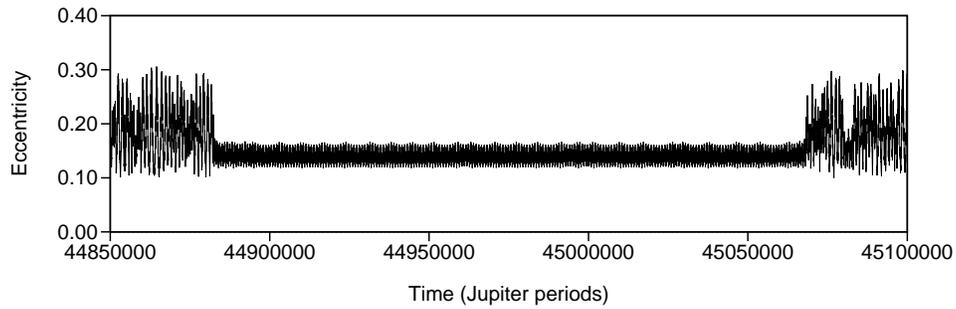}
\end{center}
\caption{The eccentricity behaviour of the same trajectory
when the winding number $Q$ is equal to one.}
\label{Q_1}
\end{figure}

In Figs.~\ref{ECC} and~\ref{Q_1}, two intermittent regimes
in variations of the eccentricity of a single asteroidal orbit
are shown. The starting values and values of parameters for
Wisdom's map~\cite{W83} are the same as cited above, except that
the semimajor axis $a_0 = 0.48088$, the eccentricity $e_0 = 0.05$,
and Jupiter's eccentricity $e_J = 0.044$. Fig.~\ref{ECC}
illustrates the well-known ``usual'' intermittent behaviour,
characterized by sporadic eccentricity bursts.
In Fig.~\ref{Q_1}, an example of transition to a regime
with both angles $\kappa$ and $\sigma$ rotating with one and
the same frequency (i.e. $Q = 1$) is shown. One can see that
this regime of the motion in the resonance $Q = 1$ only formally
can be described as a burst-like behaviour; in fact, the increase
in eccentricity from the usual quasiregular level
(compare Figs.~\ref{ECC} and~\ref{Q_1}) is relatively small.
This is because the eccentricity does not straightforwardly
correspond to the action-like variable in the approximating SM.

Can the tangency condition be used to predict emergence of
usual intermittent behaviour, that with major eccentricity
bursts? During the bursts, angles $\sigma$ and $\kappa$ librate
synchronously, i.e. a marginal integer resonance exists
at the border of the libration side of the main chaotic layer.
Therefore the emergence of marginal resonances at both sides
of the layer seems to be an approximately simultaneous event,
with respect to variation of parameters of the problem.
This implies an approximate, though not perfect one, symmetry
in the maximum relative energy deviations in the motion near the
separatrix at both sides of the layer. Therefore the orbit can
exhibit transitions to marginal resonances at both sides. What is more,
the location of these resonances should be more or less symmetric.
In the example of the exact SM presented above, these conditions are
approximately satisfied, and bursts of the relative energy emerge
almost simultaneously, with respect to variation of the parameter $W$,
both for circulations and librations.

A proof of the assertion in the considered asteroidal problem
that marginal resonances at both sides of the layer appear almost
simultaneously would provide a useful criterion
for emergence of eccentricity bursts, since
the behaviour at the rotation side of the layer satisfies
Chirikov's paradigm~\cite{C79} and on this reason
is more theoretically tractable.

\section{Conclusions}

Intermittent behaviour, namely sporadic bursts of the relative
energy from a ``quiet'' level, is prominent in the dynamics of the
separatrix map (SM) when its parameters have particular values.
Major bursts become possible when, upon variation of parameters,
a heteroclinic connection is formed between the main chaotic layer
and the narrow chaotic layer of an integer resonance, i.e. they
start to overlap; thus a marginal integer resonance emerges.
This phenomenon can be described by an approximate scheme:
the unperturbed separatrix of an integer resonance starts to be
tangent to (touches) the border of the main chaotic layer.
The condition~(\ref{tc}) for this tangency allows one
to predict the onset of intermittency upon variation of
parameters.

In order to make a straightforward comparison with numeric
integrations possible, the procedure~(\ref{ws}) of synchronization
of the SM to a unified surface of section is essential.
The choice of the surface of section farthest from the
saddle point provides a complete coverage of the phase space
of the near-separatrix motion.

Conditions upon which sporadic bursts of the relative energy
emerge in the chaotic motion described by the SM
are directly applicable to studies of the asteroidal orbits
in the 3/1 Jovian resonance. Chaotic asteroidal trajectories
displaying multiple eccentricity bursts similar to each other
provide an example of Hamiltonian intermittency. Reduction of
the usual ``quiet'' mode of this motion to the SM, performed in the
frames of the planar-elliptic problem Sun--Jupiter--asteroid,
and application of the tangency condition allows one to predict
and identify a new intermittent regime.
The motion in this regime has the winding number $Q=1$,
both the model pendulum's angle and the phase angle of perturbation
rotating with one and the same frequency. In the SM description,
it corresponds to sticking of a trajectory to the
marginal resonance $m=1$. The transition to this regime is
possible because the tangency condition~(\ref{tc}) holds.

The tangency condition at the $\kappa$-rotation side of the layer,
where Chirikov's paradigm~\cite{C79} for the SM (Eq.~(\ref{sm}))
is applicable, seems to be satisfied approximately simultaneously,
upon variation of parameters, with the same condition at the
$\kappa$-libration side, where the phenomenon of major eccentricity
bursts occurs but Chirikov's paradigm~\cite{C79} does not hold.
Therefore this simultaneity, if confirmed and explained theoretically,
would represent a useful criterion for the emergence of such bursts.

\bigskip

It is a pleasure to thank Hans Scholl for useful discussions.
This work was supported in part by the Russian
Foundation of Fundamental Research under Grant 95-02-05301-a.

\newpage

\end{document}